\documentclass[aps,prb,reprint,amsmath,superscriptaddress,amssymb]{revtex4-1}

\usepackage{graphicx, bm}
\usepackage{fp}
\usepackage[separate-uncertainty=true]{siunitx}
\usepackage{url}
\usepackage{color}
\usepackage{hyperref}
\usepackage{textcomp}
\DeclareSIUnit\permille{\text{\textperthousand}}
\newcommand{\fig}[1]{Fig.~\ref{fig:#1}}
\newcommand{\eqn}[1]{Eq.~\ref{eqn:#1}}
\renewcommand{\sec}[1]{Sec.~\ref{sec:#1}}
\newcommand{\Ms}{\ensuremath{M_\mathrm{s}}}
\newcommand{\Hani}{\ensuremath{H_\mathrm{ani}}}
\newcommand{\fres}{\ensuremath{f_\mathrm{res}}}
\newcommand{\df}{\ensuremath{\Delta f}}

\newcommand{\dfKL}{\ensuremath{\Delta f_\mathrm{KL}}}
\newcommand{\fe}{\ensuremath{f_\mathrm{res}^{110}}}
\newcommand{\fv}{\ensuremath{f_\mathrm{res}^{440}}}
\newcommand{\taure}{\ensuremath{\tau_\mathrm{RE}}}
\newcommand{\muB}{\ensuremath{\mu_\mathrm{B}}}
\newcommand{\kB}{\ensuremath{k_\mathrm{B}}}

\usepackage[sort&compress]{natbib}
\usepackage{color}

\begin{document}

\title{Temperature dependent magnetic damping of yttrium iron garnet spheres}

\author{H. Maier-Flaig}
\email{hannes.maier-flaig@wmi.badw.de}
\affiliation{Walther-Mei\ss ner-Institut, Bayerische Akademie der Wissenschaften, 85748 Garching, Germany}
\affiliation{Physik-Department, Technische Universit\"{a}t M\"{u}nchen, 85748 Garching, Germany}

\author{S. Klingler}
\affiliation{Walther-Mei\ss ner-Institut, Bayerische Akademie der Wissenschaften, 85748 Garching, Germany}
\affiliation{Physik-Department, Technische Universit\"{a}t M\"{u}nchen, 85748 Garching, Germany}

\author{C. Dubs}
\affiliation{INNOVENT e.V. Technologieentwicklung, 07745 Jena, Germany}

\author{O. Surzhenko}
\affiliation{INNOVENT e.V. Technologieentwicklung, 07745 Jena, Germany}

\author{R. Gross}
\affiliation{Walther-Mei\ss ner-Institut, Bayerische Akademie der Wissenschaften, 85748 Garching, Germany}
\affiliation{Physik-Department, Technische Universit\"{a}t M\"{u}nchen, 85748 Garching, Germany}
\affiliation{Nanosystems Initiative Munich, 80799 M\"{u}nchen, Germany}

\author{M. Weiler}
\affiliation{Walther-Mei\ss ner-Institut, Bayerische Akademie der Wissenschaften, 85748 Garching, Germany}
\affiliation{Physik-Department, Technische Universit\"{a}t M\"{u}nchen, 85748 Garching, Germany}

\author{H. Huebl}
\affiliation{Walther-Mei\ss ner-Institut, Bayerische Akademie der Wissenschaften, 85748 Garching, Germany}
\affiliation{Physik-Department, Technische Universit\"{a}t M\"{u}nchen, 85748 Garching, Germany}
\affiliation{Nanosystems Initiative Munich, 80799 M\"{u}nchen, Germany}

\author{S. T. B. Goennenwein}
\affiliation{Walther-Mei\ss ner-Institut, Bayerische Akademie der Wissenschaften, 85748 Garching, Germany}
\affiliation{Physik-Department, Technische Universit\"{a}t M\"{u}nchen, 85748 Garching, Germany}
\affiliation{Nanosystems Initiative Munich, 80799 M\"{u}nchen, Germany}
\affiliation{Institut f\"{u}r Festk\"{o}rperphysik, Technische Universit\"{a}t Dresden, 01062 Dresden, Germany}
\affiliation{Center for Transport and Devices of Emergent Materials, Technische Universit\"{a}t Dresden, 01062 Dresden, Germany}

\date{\today}

\begin{abstract}
We investigate the temperature dependent microwave absorption spectrum of an yttrium iron garnet sphere as a function of temperature (\SI{5}{\kelvin} to \SI{300}{\kelvin}) and frequency (\SI{3}{\giga\hertz} to \SI{43.5}{\giga\hertz}). At temperatures above \SI{100}{\kelvin}, the magnetic resonance linewidth increases linearly with temperature and shows a Gilbert-like linear frequency dependence. At lower temperatures, the temperature dependence of the resonance linewidth at constant external magnetic fields exhibits a characteristic peak which coincides with a non-Gilbert-like frequency dependence. The complete temperature and frequency evolution of the linewidth can be modeled by the phenomenology of slowly relaxing rare-earth impurities and either the Kasuya-LeCraw mechanism or the scattering with optical magnons. Furthermore, we extract the temperature dependence of the saturation magnetization, the magnetic anisotropy and the $g$-factor.
\end{abstract}


\maketitle

\section{Introduction}
The magnetization dynamics of the ferrimagnetic insulator yttrium iron garnet (YIG) recently gained renewed interest as YIG is considered an ideal candidate for spintronic applications as well as spin-based quantum information storage and processing\cite{Zhang2014,Zhang2015,Tabuchi2014,Bai2015} due to the exceptionally low damping of magnetic excitations as well as its magneto-optical properties\cite{Klingler2016,ViolaKusminskiy2016,Hisatomi2016}.
In particular, considerable progress has been made in implementing schemes such as coupling the magnetic moments of multiple YIG spheres\cite{Lambert2015,Zhang2015} or interfacing superconducting quantum bits with the magnetic moment of a YIG sphere\cite{Tabuchi2014,Tabuchi2015}.

Magnetization dynamics in YIG have been investigated in a large number of studies in the 1960s.\cite{Spencer1959,Sparks1960,Belov1961} However, a detailed broadband study of the magnetization dynamics in particular for low temperatures is still missing for bulk YIG.
Nevertheless, these parameters are essential for the design and optimization of spintronic and quantum devices. 
Two recent studies\cite{Haidar2015,Jermain2016} consider the temperature dependent damping of YIG thin films. \citet{Haidar2015} report a large Gilbert-like damping of unknown origin, while the low damping thin films investigated by \citet{Jermain2016} show a similar behavior as reported here.

Our systematic experiments thus provide an important link between the more recent broadband studies on YIG thin films and the mostly single-frequency studies from the 1960s:
We investigate the magnetostatic spin wave modes measured in a YIG sphere using broadband magnetic resonance up to 43.5 GHz in the temperature range from 5 to 300K. We extract the temperature dependent
magnetization, the $g$-factor and the magnetic anisotropy of YIG. Additionally, we focus our analysis on the temperature dependent damping properties of YIG and identify the phenomenology of slowly relaxing rare-earth
impurities and either the Kasuya-LeCraw mechanism or the scattering with
optical magnons as the microscopic damping mechanism.

% Structure of the paper
The paper is organized as follows. We first give a short introduction into the experimental techniques followed by a brief review  of the magnetization damping mechanisms reported for YIG. Finally, we present the measured data and compare the evolution of the linewidth with temperature and frequency with the discussed damping models. The complete set of raw data and the evaluation routines are publicly available.\cite{Maier-Flaig2017-YIG-damping-data}

\section{Experimental details and ferromagnetic resonance theory}
\label{sec:exp}
\begin{figure}
\includegraphics[width=0.45\textwidth]{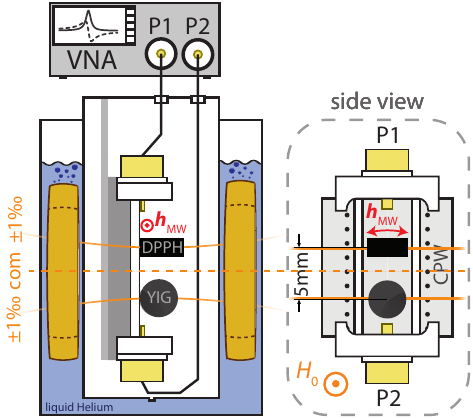}
\caption{The coplanar waveguide (CPW), on which the YIG sphere and a DPPH marker are mounted (right), is inserted into a magnet cryostat (left). 
The microwave transmission through the setup is measured phase sensitively using a vector network analyzer (VNA). As the Oersted field $h_\mathrm{MW}$ (red) around the center conductor of the CPW extends into the YIG sphere and the DPPH, we can measure the microwave response spectra of both  samples.
A superconducting magnet provides the static external magnetic field $H_0$ (orange) at the location of the sample. Also shown are the lines corresponding to the specified \SI{1}{\permille} homogeneity of the field for a on-axis deviation from the center of magnet (com).
%Obviously not to scale. Did you notice the nice gold color and the bubbles? :)
}
\label{fig:setup}
\end{figure}

%Experimental setup
The experimental setup for the investigation of the temperature dependent broadband ferromagnetic resonance (bbFMR) is shown schematically in \fig{setup}. It consists of a coplanar wave guide (CPW) onto which a \SI{300}{\micro\meter} diameter YIG sphere is mounted above the \SI{300}{\micro\meter} wide center conductor. The [111] direction of the single crystalline sphere is aligned along the CPW surface normal as confirmed using Laue diffraction (not shown).
We mount a pressed diphenylpicrylhydrazyl (DPPH) powder sample in a distance of approximately \SI{5}{\milli\meter} from the sphere. The identical sample with the same alignment has been used in Ref.~\citenum{Klingler2017}.
This assembly is mounted on a dip stick in order to place the YIG sphere in the center of a superconducting magnet (Helmholtz configuration) in a Helium gas-flow cryostat. End-launch connectors are attached to the CPW and connected to the two ports of a vector network analyzer (VNA) measuring the phase sensitive transmission of the setup up to \SI {43.5}{\giga\hertz}.

% FMR primer
The sphere is placed within the microwave Oersted field $h_\mathrm{MW}$ of the CPW's center conductor which is excited with a continuous wave microwave of variable frequency.
We apply a static external magnetic field $H_0$ perpendicular to the CPW surface and thus $h_\mathrm{MW}$ is oriented primarily perpendicular to $H_0$. The microwave Oersted field can therefore excite magnetization precession at frequencies that allow a resonant drive. The magnetization precession is detected by electromagnetic induction via the same center conductor.\cite{Kalarickal2006} This induction voltage in combination with the purely transmitted microwave signal is measured phase sensitively as the complex scattering parameter $S_{21}^\mathrm{raw}\left(\omega\right)$ at port 2 of the VNA.

% Background subtraction method
The frequency-dependent background is eliminated as follows: A static external magnetic field sufficiently large that no resonances are expected in the given microwave frequency range is applied and the transmission at this field is recorded as the background reference $S_{21}^\mathrm{BG}$. Then, the external field is set to the value at which we expect resonances of YIG in the given frequency range and record the transmission $S_{21}^\mathrm{raw}$. We finally divide $S_{21}^\mathrm{raw}$ by $S_{21}^\mathrm{BG}$. 
This corrects for the frequency dependent attenuation and the electrical length of the setup. We choose this background removal method over a microwave calibration because it additionally eliminates the field and temperature dependence of $S_{21}$ that arises from the thermal contraction and movement of the setup and magnetic materials in the microwave connectors. 
In the following we always display $S_{21} = S_{21}^\mathrm{raw}/S_{21}^\mathrm{BG}$.

% FMR susceptibility & fit function
For the evaluation of the magnetization dynamics, we fit the transmission data to $S_{21}= -i f Z \chi+A_1+A_2 f$  for each $H_0$. Here,  $A_{1,2}$ describe a complex-valued background and
\begin{equation}
  \label{eqn:polder}
  \chi\left(f, H_0\right) =  \frac{
  		\mu_0 \Ms \frac{\gamma}{2\pi} \left(\frac{\gamma}{2\pi} \mu_0 H_0 - i \Delta f \right)
  	}
  	{
  		\fres^2  - f^2 - i f \Delta f
  	}
\end{equation}
is the ferromagnetic high-frequency susceptibility.\cite{Schneider2007,Kalarickal2006}
The free parameters of the fit are the resonance frequency \fres, the full width at half maximum (FWHM) linewidth \df\ as well as the complex scaling parameter $Z$, which is proportional to the strength of the inductive coupling between the specific magnetic resonance mode and the CPW. For a given fixed magnetic field the fit parameters $\frac{\gamma}{2\pi}$ (gyromagnetic ratio) and $\Ms$ (saturation magnetization) are completely correlated with $Z$ and are thus fixed. They are later determined from fitting the dispersion curves.\cite{Nembach2011}

% FMR dispersion and modes
In spheres various so-called magnetostatic modes (MSM) arise due to the electromagnetic boundary conditions.\cite{Roschmann1977} These modes can be derived from the Landau-Lifshitz equation in the magnetostatic limit ($\vec{\nabla} \times \vec{H} = 0$) for insulators.\cite{Walker1957} The lineshape of all modes is given by \eqn{polder}. 
Due to the different spatial mode profiles and the inhomogeneous microwave field, the inductive coupling and thus $Z$ is mode dependent.%
\footnote{
  Due to the comparable dimensions of center conductor width and sphere diameter, we expect that the sphere experiences an inhomogeneous microwave magnetic field with its main component parallel to the surface of the CPW and perpendicular to its center conductor.
  As the microwave magnetic field is sufficiently small to not cause any non-linear effects, a mode dependent excitation efficiency is the only effect of the microwave magnetic field inhomogeneity.
}
A detailed review of possible modes, their distribution and dispersion is given in \citet{Roschmann1977}. We will only discuss the modes (110) and (440) in detail in the following as all the relevant characteristics of all other modes can be related to these two modes. 
Their linear dispersions are given by\cite{Roschmann1977}
\begin{equation}
  \label{eqn:110}
  \fe = \frac{\gamma}{2\pi} \mu_0\left( H_0 +  \Hani\right)
\end{equation}
\begin{equation}
  \label{eqn:440}
  \fv =  \frac{\gamma}{2\pi} \mu_0 \left( H_0 +  \Hani + \frac{\Ms}{9}\right)
\end{equation}
where $H_\mathrm{ani}$ is the magnetic anisotropy field and $\frac{\gamma}{2\pi}$ is the gyromagnetic ratio which relates to the $g$-factor by $\frac{\gamma}{2\pi} = \frac{\muB}{h}g$. It is thus generally assumed that $g$ is the same for all modes. We note that the apparent $g$-factor may still vary in between modes if the modes experience a different anisotropy.\cite{Kittel1949,Bruno1989} Such an anisotropy contribution can be caused by surface pit scattering as it affects modes that are localized at the surface stronger than bulk like modes\cite{Nemarich1964}. In our experiment, no such variation in $g$ coinciding with a change in anisotropy was observed and we use a mode number independent $g$ in the following.

Knowledge of the dispersion relations of the two modes allows to determine the saturation magnetization from
\begin{equation}
        \label{eqn:Ms}
        \mu_0 \Ms\left(T\right) = 9 \frac{2\pi}{\gamma} \Delta f_\mathrm{M} = 9 \frac{2\pi}{\gamma} \left (\fv - \fe\right).
\end{equation}
The anisotropy field is extracted by extrapolating the dispersion relations in Eqs.~\eqref{eqn:110} and \eqref{eqn:440}  to $H_0 = 0$.

The temperature dependent linewidth \df\ of the modes is the central result of this work. For a short review of the relevant relaxation processes we refer to the dedicated \sec{relaxation}.

In this work, we investigate the $T$-dependence of \Ms, \Hani, $g$ and \df. 
Accurate determination of the $g$-factor and the anisotropy $H_\mathrm{ani}$ requires accurate knowledge of $H_0$.
In order to control the  temperature of the YIG sphere and CPW, they are placed in a gas-flow cryostat as displayed schematically in \fig{setup}.
The challenge in this type of setup is the exact and independent determination of the static magnetic field and its spatial inhomogeneity.
Lacking an independent measure of $H_0$,%
\footnote{
  The resonance frequency of the DPPH sample that has been measured simultaneously was intended as a field calibration but can not be utilized due to the magnetic field inhomogeneity. In particular, since the homogeneity of our superconducting magnet system is specified to \SI{1}{\permille} for an off-axis deviation of \SI{2.5}{\milli\meter}, the spatial separation of \SI{5}{\milli\meter} of the DPPH and the YIG sphere already falsifies DPPH as an independent magnetic field standard. Placing DPPH and YIG in closer proximity is problematic as the stray field of the YIG sphere will affect the resonance frequency of the DPPH. Note further that we are not aware of any reports showing the temperature independence of the DPPH $g$-factor with the required accuracy.
}
we only report the relative change of $g$ and $H_\mathrm{ani}$ from their respective room temperature values which were determined separately using the same  YIG sphere.\cite{Klingler2017} Note that we determine the resonance frequencies directly in frequency space. Our results on linewidth and magnetization are hence independent of a potential  uncertainty in the absolute magnitude of $H_0$ and its inhomogeneity.

\section{Relaxation theory}
\label{sec:relaxation}

When relaxation properties of ferromagnets are discussed today, the most widely applied model is the so-called Gilbert type damping. This purely phenomenological model is expressed in a damping term of the form $\alpha M \times \frac{\mathrm{d}M}{\mathrm{d}t}$ in the Landau-Lifshitz equation. It describes a viscous damping, i.e. a resonance linewidth that depends linearly on the frequency. A linear frequency dependence is often found in experiments and the Gilbert damping parameter $\alpha$ serves as a figure of merit of the ferromagnetic damping that allows to compare samples and materials. It contains, however, no insight into the underlying physical mechanisms.

In order to understand the underlying microscopic relaxation processes of YIG, extensive work has been carried out. Improvements on both the experimental side (low temperatures\cite{Dillon1958}, temperature dependence \cite{Kasuya1961,Spencer1959,Spencer1961}, separate measurements of $M_{z}$ and $M_{xy}$\cite{Sparks1960}) and on the sample preparation (varying the surface pit size\cite{Nemarich1964}, purifying Yttrium\cite{Spencer1959}, doping YIG with silicon\cite{Sparks1964} and rare-earth elements \cite{Sparks1964,Clarke1965,Clarke1965a,Sparks1967}) led to a better understanding of these mechanisms.

However, despite these efforts the microscopic origin of the dominant relaxation mechanism for bulk YIG at room temperature is still under debate. It has been described by a two-magnon process by \citet{Kasuya1961}~(1961).
In this process, a uniformly-precessing magnon ($k=0$) relaxes under absorption of a phonon to a $k\neq0$ magnon. If the thermal energy $\kB T$ is much larger than the energy of the involved magnons and phonons ($T>\SI{100}{\kelvin}$) and low enough that no higher-order processes such as four-magnon scattering play a role ($T<\SI{350}{\kelvin}$), the Kasuya-LeCraw process yields a linewidth that is linear in frequency and temperature: $\dfKL \propto T,f$.\cite{Kasuya1961,Sparks1964}
This microscopic process is therefore considered to be the physical process that explains the phenomenological Gilbert damping for low-damping bulk YIG. More recently, \citet{Cherepanov1993} pointed out that the calculations by \citet{Kasuya1961} assume a quadratic magnon dispersion in $k$-space which is only correct for very small wave numbers $k$. Taking into account a more realistic magnon dispersion  (quadratic at low $k$, linear to higher $k$), the Kasuya-LeCraw mechanism gives a value for the relaxation rate that is not in line with the experimental results. Cherepanov therefore developed an alternative model that traces back the linear frequency and temperature dependence at high temperatures ($\SI{150}{\kelvin}$ to $\SI{300}{\kelvin}$) to the interaction of the uniform-precession mode with optical magnons of high frequency. Recently, atomistic calculations by \citet{Barker2016} confirmed the assumptions on the magnon spectrum that are necessary for the quantitative agreement of the latter theory with experiment.

Both theories, the Kasuya-LeCraw theory and the Cherepanov theory, aim to describe the microscopic origin of the intrinsic damping. They deviate in their prediction only in the low-temperature ($T<\SI{100}{\kelvin}$) behavior.\cite{Sparks1964} 
At these temperatures, however, impurities typically dominate the relaxation and mask the contribution of the intrinsic damping process. 
Therefore, the dominant microscopic origin of the YIG damping at temperatures above 150K has not been unambiguously determined to date.

If rare-earth impurities with large orbital momentum exist in the crystal lattice, their exchange coupling with the iron ions introduces an additional relaxation channel  for the uniform precession mode of YIG. Depending on the relaxation rate of the rare-earth impurities with respect to the magneto--dynamics of YIG, they are classified into slowly and fast relaxing rare-earth impurities. This is an important distinction as the efficiency of the relaxation of the fundamental mode of YIG via the rare-earth ion to the lattice at a given frequency depends on the relaxation rate of the rare-earth ion and the strength of the exchange coupling.   In both the slow and the fast relaxor case, a characteristic peak-like maximum is observed in the linewidth vs. temperature dependence at a characteristic, frequency-dependent temperature\cite{Belov1961}. The frequency dependence of this peak temperature allows to distinguish fast and slowly relaxing rare-earth ions: The model of a fast relaxing impurity predicts that the peak temperature is constant, while in the case of slowly relaxing rare-earth ions the peak  temperature is experted to increase with increasing magnetic field (or frequency). 
The relaxation rate of rare-earths $\taure$ is typically modeled by a direct magnon to phonon relaxation, an Orbach processes\cite{Orbach1961,Clarke1965b} that involves two phonons, or a combination of both. The inverse relaxation rate of an Orbach process is described by $\frac{1}{\tau^\mathrm{Orbach}} = \frac{B}{e^{\Delta/(\kB T)} - 1}$ with the crystal field splitting $\Delta$ and a proportionality factor $B$. A direct process leads to an inverse relaxation rate of $\frac{1}{\tau^\mathrm{direct}} = \frac{1}{\tau_0}  \coth{\frac{\delta}{2 \kB T}} 
$ with $\tau_0$, the relaxation time at $T=\SI{0}{\kelvin}$.
It has been found experimentally that most rare-earth impurities are to be classified as slow relaxors.\cite{Sparks1964} The sample investigated here is not intentionally doped with a certain rare-earth element and the peak frequency and temperature dependence indicates a slow relaxor. We therefore focus on the slow relaxing rare-earth impurity model in the following.

Deriving the theory of the slowly relaxing impurities has been performed comprehensively elsewhere.\cite{Sparks1964} The linewidth contribution caused by a slowly relaxing rare-earth impurity is given by\cite{Clarke1965}:
\begin{equation}
  \label{eqn:SR}
  \df^\mathrm{SR} = \frac{C}{2\pi} \frac{f\taure}{1+\left(f \taure\right)^2}
\end{equation}
with $C\propto \frac{1}{\kB T} \operatorname{sech}\left(\frac{\delta_\mathrm{a}}{2\kB T}\right)$. Therein, $\delta_\mathrm{a}$ is the splitting of the rare-earth Kramers doublet which is given by the temperature independent exchange interaction between the iron ions and the rare-earth ions.

Also Fe$^{2+}$ impurities in YIG give rise to a process that leads to a linewidth peak at a certain temperature. The physical origin of this so-called valence exchange or charge-transfer linewidth broadening is electron hopping between the iron ions.\cite{Sparks1964} Simplified, it can be viewed as a two level system just like a rare-earth ion and thus results in the same characteristic linewidth maximum as a slowly relaxing rare-earth ion. For valence exchange, the energy barrier $\Delta_\mathrm{hop}$ that needs to be overcome for hopping determines the time scale of the process. The two processes, valence exchange and rare-earth impurity relaxation, can therefore typically not be told apart from FMR measurements only. In the following, we use the slow relaxor mechanism exclusively. This model consistently describes our measurement data and the resulting model parameters are in good agreement with literature. We would like to emphasize, however, that the valence exchange mechanism as the relevant microscopic process resulting for magnetization damping can not be ruled out from our measurements.

\section{Experimental results and discussion}
\label{sec:discuss}
\begin{figure}
\includegraphics[width=0.45\textwidth]{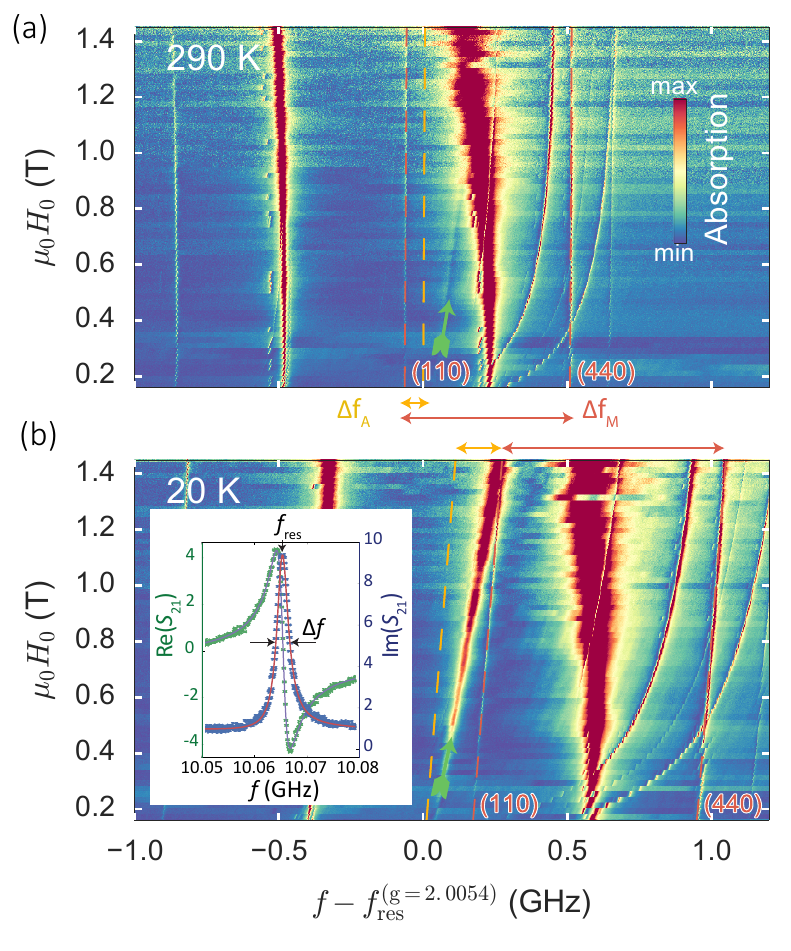}
\caption{
Eigenmode spectra of the YIG sphere at (a) \SI{290}{\kelvin} and (b) \SI{20}{\kelvin}.
The (110) and (440) MSM are marked with red dashed lines.
The change in their slope gives the change of the $g$-factor of YIG.
Their splitting ($\Delta f_\mathrm{M}$, red arrow) depends linearly on the YIG magnetization.
The increase in $\Ms$ to lower temperatures is already apparent from the
increased splitting $\Delta f_\mathrm{M}$.
Marked in orange is the offset of the resonance frequency $\Delta f_\mathrm{A}$
extrapolated to $H_0=0$ resulting from anisotropy fields $\Hani$ present in the sphere.
The green marker denotes the position of the DPPH resonance line which increases
in amplitude considerably to lower temperatures.
Inset: $S_{21}$ parameter (data points) and fit (lines) at $\mu_0 H = \SI{321}{\milli\tesla}$
and $T=\SI{20}{\kelvin}$.
}
\label{fig:spectra}
\end{figure}

%% #### Spectra
Two exemplary $S_{21}$ broadband spectra recorded at two distinct temperatures are shown in \fig{spectra}. The color-coded magnitude $\left| S_{21} \right|$ is a measure for the absorbed microwave power. High absorption (bright color) indicates the resonant excitation of a MSM in the YIG sphere or the excitation of the electron paramagnetic resonance of the DPPH. In the color plot the color scale is truncated in order to improve visibility of small amplitude resonances. In addition, the frequency axis is shifted relative to the resonance frequency of a linear dispersion with $g=2.0054$ ($f_\mathrm{res}^{g=2.0054}=\frac{g \muB}{h} \mu_0 H$) for each field. In this way, modes with $g=2.0054$ appear as vertical lines. 
A deviating $g$-factor is therefore easily visible as a different slope.
Comparing the spectra at \SI{290}{\kelvin} [\fig{spectra}\,(a)] to the spectra at \SI{20}{K} [\fig{spectra}\,(b)], an increase of the $g$-factor is observed for all resonance modes upon reducing the temperature.
The rich mode spectrum makes it necessary to carefully identify the modes and assign mode numbers. Note that the occurrence of a particular mode in the spectrum depends on the position of the sphere with respect to the CPW due to its inhomogeneous excitation field. We employ the same method of identifying the modes as used in Ref.~\citenum{Klingler2017} and find consistent mode spectra. As mentioned before, we do not use the DPPH resonance (green arrow in \fig{spectra}) but the (110) YIG mode as field reference. For this field reference, we take $g(\SI{290}{\kelvin}) = 2.0054$ and $\frac{\gamma}{2\pi} \mu_0 H_\mathrm{ani}(\SI{290}{\kelvin}) = \SI{68.5}{\mega\hertz}$ determined for the same YIG sphere at room temperature in an electromagnet with more accurate knowledge of the applied external magnetic field.\cite{Klingler2017} 
The discrepancy of the DPPH $g$-value from the literature values of $g=2.0036$ is attributed to the non-optimal location of the DPPH specimen in the homogeneous region of the superconducting magnet coils.

In \fig{spectra}, the fitted dispersion of the (110) and (440) modes are shown as dashed red lines. As noted previously, we only analyze these two modes in detail as all parameters can be extracted from just two modes. The (110) and (440) mode can be easily and unambiguously identified by simply comparing the spectra with the ones found in Ref~\citenum{Klingler2017}. Furthermore, at high fields, both modes are clearly separated from other modes. 
This is necessary as modes can start hybridizing when their (unperturbed) resonance frequencies are very similar (cf. low-field region of \fig{spectra}~(b)) which makes a reliable determination of the linewidth and resonance frequency impossible. 
These attributes make the (110) and the (440) mode the ideal choice for the analysis.

\begin{figure}
\includegraphics[width=0.45\textwidth]{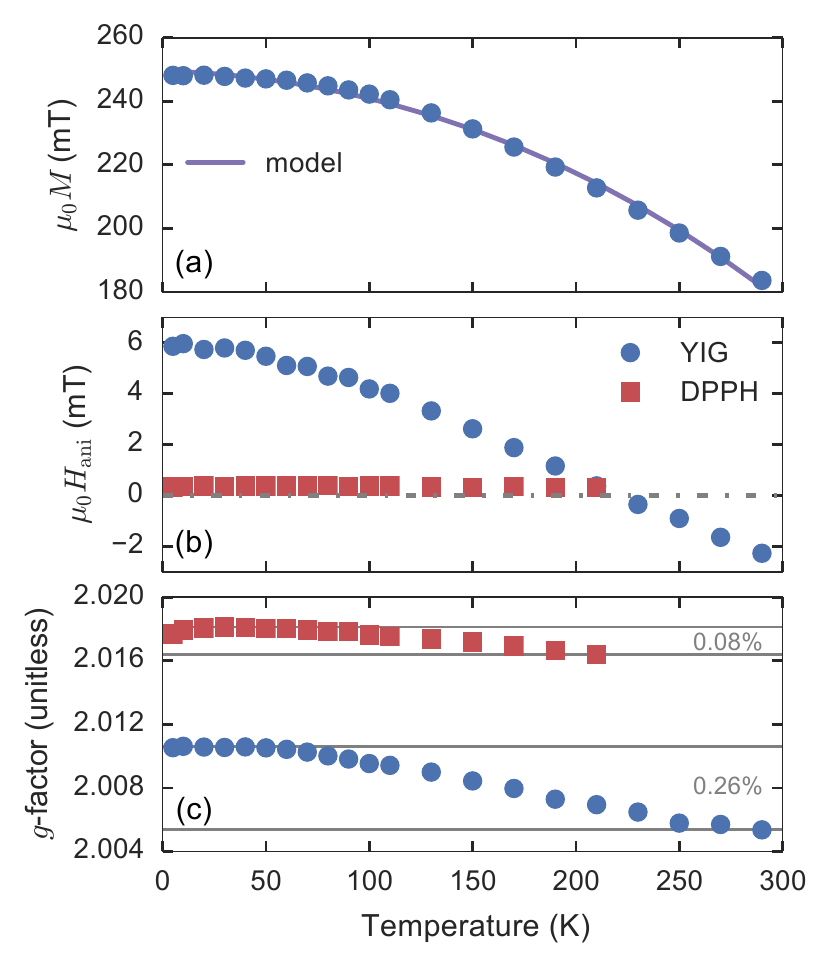}
\caption{
\textbf{(a)} YIG magnetization as function of temperature extracted from the (110) and (440) mode dispersions using \eqn{Ms}. The purple line shows the fit to a Bloch model (cf. parameters in the main text).
\textbf{(b)} YIG anisotropy field $\mu_0 H_\mathrm{ani}\left(T\right) = \frac{2\pi}{\gamma} \Delta f_\mathrm{ani}$. Red squares: Same procedure applied to the DPPH dispersion as reference.
\textbf{(c)} YIG $g$-factor (blue circles). 
For reference, the extracted DPPH $g$-factor is also shown (red squares).
The gray numbers indicate the relative change of the $g$-factors from the lowest to the highest measured temperature (gray horizontal lines).
As we use the YIG (110) mode as the magnetic field reference, the extracted value of $g$ and $\Hani$ at \SI{300}{\kelvin} are fixed to the values determined in the room temperature setup.\cite{Klingler2017} 
}
\label{fig:magnetization}
\end{figure}

%% #### Magnetization
As described in \sec{exp}, we simultaneously fit the (110) and the (440) dispersions with the same $g$-factor in order to extract $\Ms$, $\Hani$ and $g$. In the fit, we only take the high-field dispersion of the modes into account where no other modes intersect the dispersion of the (110) and (440) modes. The results are shown in \fig{magnetization}. Note that the statistical uncertainty from the fit is not visible on the scale of any of the parameters $\Ms$, $\Hani$ and $g$.
Following the work of \citet{Solt1962}, we model the resulting temperature dependence of the magnetization (\fig{magnetization}~(a)) with the Bloch-law taking only the first order correction into account:
\begin{equation}
  M_\mathrm{s} = M_0\left(1-aT^{\frac{3}{2}} - bT^{\frac{5}{2}}\right).
\end{equation}
The best fit is obtained for $\mu_0 M_0 = \SI[separate-uncertainty=false]{249.5(5)}{\milli\tesla}$,
$a=\SI{23(3)e-6}{\kelvin^{-3/2}}$ and
$b=\SI{1.08(11)e-7}{\kelvin^{-5/2}}$. 
The obtained fit parameters depend strongly on the temperature window in which the data is fitted.
Hence, the underlying physics determining the constants $a$ and $b$ cannot be resolved.%
\footnote{
  Note that we failed to reproduce the fit of Ref.~\citenum{Solt1962} using the data provided in this paper and that the reasonable agreement with the there-reported fit parameters might be coincidence.
}
Nevertheless, the temperature dependence of $\Ms$ is in good agreement with the results determined using a vibrating sample magnetometer.\cite{Elmer1964} 

In particular, also the room temperature saturation magnetization of $\mu_0 \Ms(\SI{300}{\kelvin}) = \SI{180\pm0.8}{\milli\tesla}$ is in perfect agreement  with values reported in literature.\cite{Hansen1974,Winkler1981}
Note that the splitting of the modes is purely in frequency space and thus errors in the field do not add to the uncertainty. 
We detect a small non-linearity of the (110) and (440) mode dispersions that is most likely due to deviations from an ideal spherical shape or strain due to the YIG mounting. 
This results in a systematic, temperature independent residual of the linear fits to these dispersions. This resulting systematic error of the magnetization is incorporated in the uncertainty given above. 
However, a deviation from the ideal spherical shape, strain in the holder or a misalignment of the static magnetic field can also modify the splitting of the modes and hence result in a different $\Ms$.\cite{White1960}
This fact may explain the small discrepancy of the value determined here and the value determined for the same sphere in a different setup at room temperature.\cite{Klingler2017}

%% #### Anisotropy
From the same fit that we use to determine the magnetization, we can deduce the temperature dependence of the anisotropy field $\mu_0 \Hani$ [\fig{magnetization}~(b)]. Most notably, \Hani\ changes sign at \SI{200}{\kelvin} which has not been observed in literature before and can be an indication that the sample is slightly strained in the holder. The resonance frequency of DPPH extrapolated to $\mu_0 H_0=0$ ($\Delta f_\mathrm{ani}$, red squares) confirms that the error in the determined value $\Hani$ is indeed temperature independent and very close to zero. Thus, the extracted value for the anisotropy is not merely given by an offset in the static magnetic field.

%% #### g-factor
The evolution of the $g$-factor with temperature is shown in \fig{magnetization}~(c). It changes from $2.005$ at room temperature to $2.010$ at \SI{10}{\kelvin} where it seems to approach a constant value. As mentioned before, the modes' dispersion is slightly non-linear giving rise to a systematic, temperature independent uncertainty in the determination of $g$ of $\pm 0.0008$. 
The $g$-factor of YIG has been determined using the MSMs of a sphere for a few selected temperatures before.\cite{Belov1961}
Comparing our data to these results, one finds that the trend of the temperature dependence of $g$ agrees. 
However, the absolute value of $g$ and the magnitude of the variation differ.
At the same time, we find a change of the $g$-factor of DPPH that is on the scale of $0.0012$. This may be attributed to a movement of the sample slightly away from the center of magnet with changing temperature due to thermal contraction of the dip stick. In this case, the YIG $g$-factor has to be corrected by this change. The magnitude of this effect on the YIG $g$-factor can not be estimated reliably from the change of the DPPH $g$-factor alone. 
Furthermore, the temperature dependence of the DPPH $g$-factor has not been investigated with the required accuracy in literature to date to allow excluding a temperature dependence of the $g$-factor of DPPH. We therefore do not present the corrected data but conclude that we observe a change in the YIG $g$-factor from room temperature to \SI{10}{\kelvin} of at least 0.2\,\%.

%% #### Damping
Next, we turn to the analysis of the damping properties of YIG. We will almost exclusively discuss the damping of the (110) mode in the following but the results also hold quantitatively and qualitatively for the other modes.\cite{Klingler2017} 
Varying the applied microwave excitation power $P$ (not shown) confirms that no nonlinear effects such as a power broadening of the modes are observed with $P=\SI{0.1}{\milli\watt}$. 
Note that due to the microwave attenuation in the microwave cabling, the microwave field at the sample location decreases with increasing frequency for the constant excitation power.

\begin{figure*}
\includegraphics[width=0.45\textwidth]{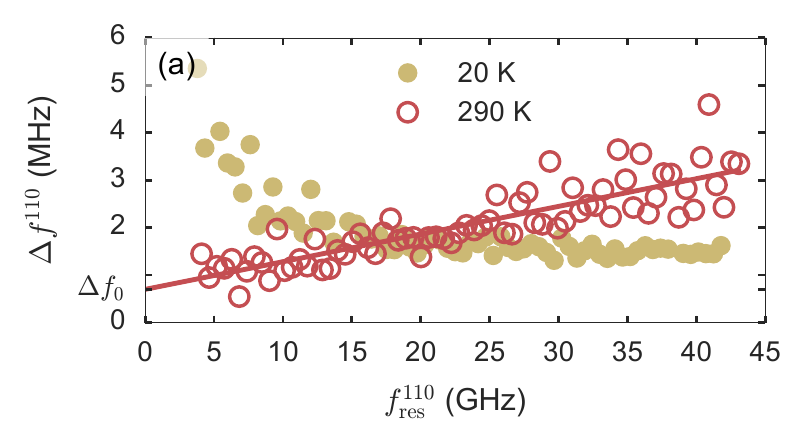}
\includegraphics[width=0.45\textwidth]{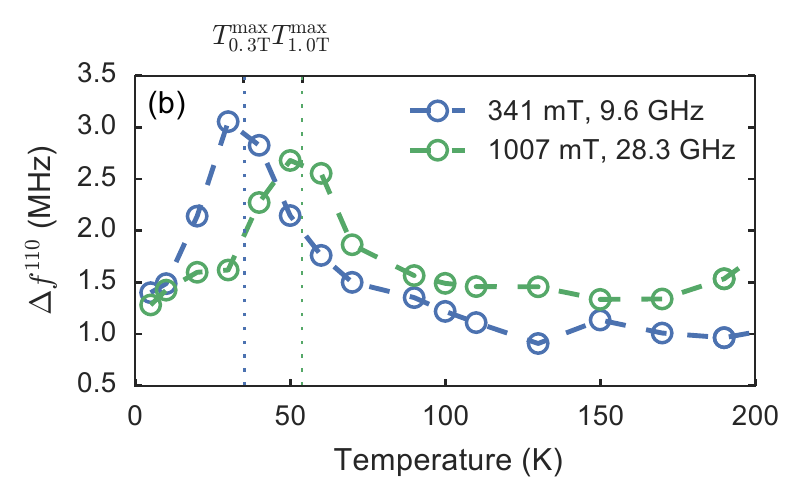}
\caption{
\textbf{(a)} Full width at half maximum (FWHM) linewidth $\Delta\fe$ of the (110) mode as a function of frequency for different temperatures. A linear Gilbert-like interpretation is justified in the high-$T$ case ($T>\SI{100}{\kelvin}$) only. Below \SI{100}{\kelvin}, the slope of $\df^{110}(\fe)$ is not linear so that a Gilbert type interpretation is no longer applicable.
\textbf{(b)} FWHM linewidth as a function of temperature for two different fixed external magnetic fields. The linewidth peaks at a magnetic field dependent temperature that can be modeled using the phenomenology of rare-earth impurities resulting in $T_\mathrm{max}$ (vertical dotted lines).}
\label{fig:peak}
\end{figure*}

\begin{figure*}
\includegraphics[width=0.45\textwidth]{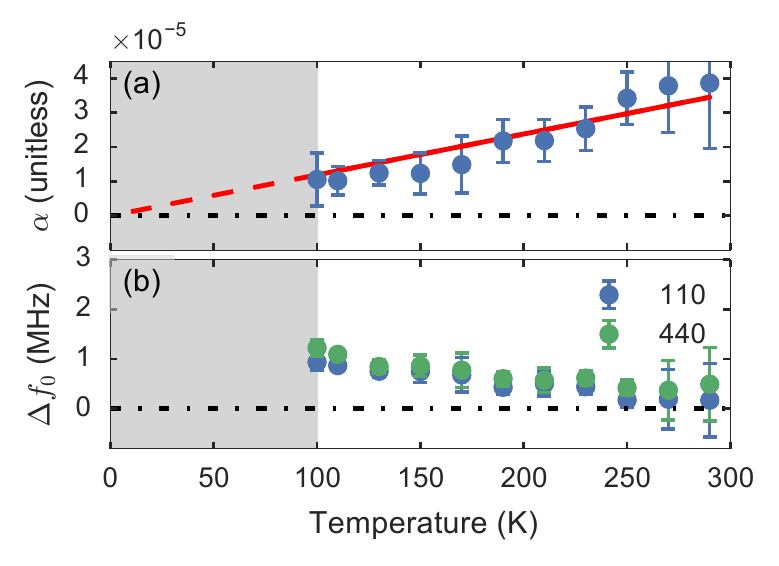}
\caption{
\textbf{(a)} Gilbert damping parameter $\alpha$ determined from the slope of a linear fit to the $\df(f,T)$ data for frequencies above \SI{20}{\giga\hertz}. The red line shows the linear dependence of the linewidth with temperature expected from the Kasuya-LeCraw process.
\textbf{(b)} Inhomogeneous linewidth $\df_0$ (intersect of the aforementioned fit with $\fe=0$) as a function of temperature. The inhomogeneous linewidth shows a slight increase with decreasing temperature down to \SI{100}{\kelvin}.
In the region where the slow relaxor dominates the linewidth (gray shaded area, cf. \fig{peak}), the linear fit is not applicable and unphysical damping parameters and inhomogeneous linewidths are extracted. 
}
\label{fig:gilbert}
\end{figure*}

% Frequency dependence of the linewidth
First, we evaluate the frequency dependent linewidth for several selected temperatures [\fig{peak}~(a)]. 
At temperatures above \SI{100}{\kelvin}, a linear dependence of the linewidth with the resonance frequency is  observed. This dependence is the usual so-called Gilbert-like damping and the slope is described by the Gilbert damping parameter $\alpha$. A linear frequency dependence of the damping in bulk YIG has been described by the theory developed by \citet{Kasuya1961} and the theory developed by \citet{Cherepanov1993} (cf. \sec{relaxation}). We extract $\alpha$ from a global fit of a linear model to the (110) and (440) linewidth with separate parameters for the inhomogeneous linewidths $\df_0^{110}$ and $\df_0^{440}$ and a shared Gilbert damping parameter $\alpha$ for all modes:\cite{Klingler2017} 
\begin{equation}
	\df = 2 \alpha f + \df_0^{110,440}
\end{equation}
The fit is shown exemplarily for the \SI{290}{\kelvin} (red) data in \fig{peak}~(a).

% Gilbert damping discussion
The Gilbert damping parameter $\alpha$ extracted using this fitting routine for each temperature is shown in \fig{gilbert}~(a). 
Consistently with both theories, $\alpha$ increases with increasing temperature. The error bars in the figure correspond to the maximal deviation of $\alpha$ extracted from separate fits for each mode. 
They therefore give a measure of how $\alpha$ scatters in between modes. 
The statistical error of the fit (typically $\pm 0.00001$) is not visible on this scale.
The Gilbert damping parameter $\alpha$ linearly extrapolated to zero temperature vanishes. Note that this is consistent with the magnon-phonon process described by \citet{Kasuya1961} but not with the theory developed by \citet{Cherepanov1993}. For room temperature, we extract a Gilbert damping of $4\times 10^{-5}$ which is in excellent agreement with the literature value.\cite{Roschmann1983,Klingler2017}
From the fit, we also extract the inhomogeneous linewidth $\df_0$, which we primarily associate with surface pit scattering (\sec{relaxation}, Ref~\citenum{Klingler2017}). In the data, a slight increase of $\df_0$ towards lower temperatures is present [\fig{gilbert}~(b)]. Such a change in the inhomogeneous linewidth can be caused by a change in the surface pit scattering contribution when the spin-wave manifold changes with $\Ms$.\cite{Klingler2017,Nemarich1964}

Note that according to \fig{gilbert}~(b) $\df_0$ is higher for the (440) mode than for the (110) mode. This is in agreement with the theoretical expectation that surface pit scattering has a higher impact on $\df_0$ for modes that are more localized at the surface of the sphere like the (440) mode compared to the more bulk-like modes such as the (110) mode\cite{Nemarich1964}.\footnote{In comparison to \citet{Klingler2017}, here, we do not see an increased inhomogeneous linewidth of the (110) mode and no secondary mode that is almost degenerate with the (110) mode. The difference can be explained by the orientation of the sphere which is very difficult to reproduce very accurately ($<\ang{1}$) between the experimental setups: The change in orientation either separates the mode that is almost degenerate to the (110) mode or makes the degeneracy more perfect in our setup. The different placement of the sphere on the CPW can also lead to a situation where the degenerate mode is not excited and therefore does not interfere with the fit.}

Turning back to \fig{peak}~(a), for low temperatures (\SI{20}{\kelvin}, blue data points), a Gilbert-like damping model is obviously not appropriate as the linewidth increases considerably towards lower frequencies instead of increasing linearly with increasing frequency. Typically, one assumes that the damping at low frequencies is dominated by so-called low field losses that may arise due to domain formation. The usual approach is then to fit a linear trend to the high-frequency behavior only. Note, however, that even though the frequency range we use is already larger than usually reported\cite{Haidar2015,Sun2012,Jermain2016}, this approach yields an unphysical, negative damping. We conclude that the model of a Gilbert-like damping is only valid for temperatures exceeding \SI{100}{\kelvin} (\fig{gilbert}) for the employed field and frequency range.

The linewidth data available in literature are typically taken at a fixed frequency and the linewidth is displayed as a function of temperature\cite{Spencer1961,Sparks1964,Belov1961}. We can approximately reproduce these results by plotting the measured linewidth at fixed $H_0$ as a function of temperature [\fig{peak}~(b)].\footnote{Naturally, the resonance frequency varies slightly ($\pm \SI{0.9}{\giga\hertz}$) between the data points because the magnetization and the anisotropy changes with temperature.} A peak-like maximum of the linewidth below \SI{100}{\kelvin} is clearly visible. For increasing magnetic field (frequency), the peak position shifts to higher temperatures. This is the signature of a slowly relaxing rare-earth impurity (\sec{relaxation}). A fast relaxing impurity is expected to result in a field-independent linewidth vs. temperature peak and can thus be ruled out.
At the peak position, the linewidth shows an increase by \SI{2.5}{\mega\hertz} which translates with the gyromagnetic ratio to a field linewidth increase of \SI{0.08}{\milli\tesla}.
For 0.1 at.\,\% Terbium doped YIG, a linewidth increase of \SI{80}{\milli\tesla} has been observed\cite{Dillon1959}. Considering that the linewidth broadening is proportional to the impurity concentration and taking the specified purity of the source material of 99.9999\% used to grow the YIG sphere investigated here, we estimate an increase of the linewidth of \SI{0.08}{\milli\tesla}, in excellent agreement with the observed value.

\begin{figure*}
\includegraphics[width=0.9\textwidth]{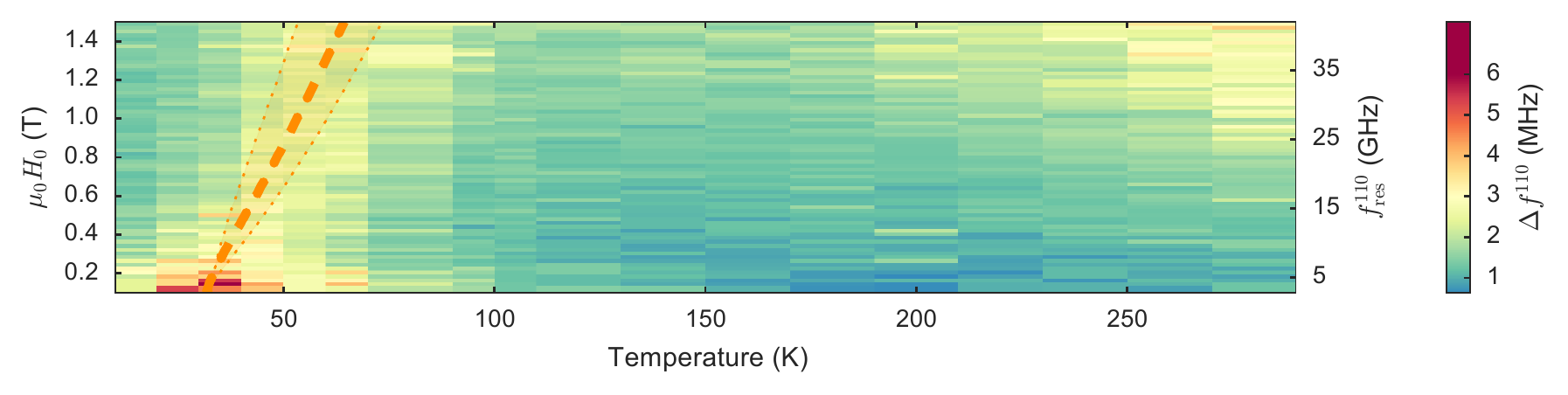}
\caption{Full map of the FWHM linewidth of the (110) mode as function of temperature and field resp. resonance frequency $\fe$. At low temperatures, only the slow relaxor peak is visible while at high temperatures the Gilbert-like damping becomes dominant. 
The position of the peak in the linewidth modeled by a slow relaxor is shown as dashed orange line. The model parameters are taken from \citet{Clarke1965} and taking $\delta_a = \SI{2.50}{\milli\electronvolt}$. The dotted lines indicate the deviation of the model for $0.5 \delta_a$ (lower $T^\mathrm{max}$) and $2 \delta_a$ (higher $T^\mathrm{max}$).
}
\label{fig:colorplot}
\end{figure*}

Modeling the linewidth data is more challenging: The model of a slowly relaxing rare-earth ion contains the exchange coupling of the rare-earth ion and the iron sublattice, and its temperature dependent relaxation frequency as parameters. As noted before, typically a direct and an Orbach process model the relaxation rate, and both of these processes have two free parameters. Unless these parameters are known from other experiments for the specific impurity and its concentration in the sample, fitting the model to the temperature behavior of the linewidth at just one fixed frequency gives ambiguous parameters. In principle, frequency resolved experiments as presented in this work make the determination of the parameters more robust as the mechanism responsible for the rare-earth relaxation is expected not to vary as a function of frequency. The complete frequency and field dependence of the linewidth is shown in \fig{colorplot}. At temperatures above approx \SI{100}{\kelvin}, the linewidth increases monotonically with field, in agreement with a dominantly Gilbert-like damping mechanisms, which becomes stronger for higher temperatures. On the same linear color scale, the linewidth peak below \SI{100}{\kelvin} and its frequency evolution is apparent. \fig{peak}~(b) corresponds to horizontal cuts of the data in \fig{colorplot} at $\mu_0 H_0=341$ and $\SI{1007}{\milli\tesla}$.

For typical YIG spheres, that are not specifically enriched with only one rare-earth element, the composition of the impurities is unknown. Different rare-earth ions contribute almost additively to the linewidth and have their own characteristic temperature dependent relaxation frequency respectively peak position. This is most probably the case for the YIG sphere of this study. The constant magnitude of the peak above \SI{0.3}{\tesla} and the constant peak width indicates that fast relaxing rare-earth ions play a minor role.
The evolution of the linewidth with $H_0$ and $f$ can therefore not be fitted to one set of parameters. We thus take a different approach and model just the shift of the peak position in frequency and temperature as originating from a single slowly relaxing rare-earth impurity. For this, we use a value for the exchange coupling energy  between the rare-earth ions and the iron sublattice in a range compatible with literature\cite{Sparks1964} of $\delta_\mathrm{a}=\SI{2.50}{\milli\electronvolt}$. To model the rare-earth relaxation rate as a function of temperature, we use the values determined by \citet{Clarke1965a} for  Neodymium doped YIG: $\tau_0=\SI{2.5e-11}{\second}$ for the direct process and $\Delta=\SI{10.54}{\milli\electronvolt}$ and $B=\SI{9e11}{\per\second}$ for the Orbach process. The model result, i.e. the peak position, is shown as dashed line in \fig{colorplot} and shows good agreement with the data.
This indicates that, even though valence exchange and other types of impurities cannot be rigourously excluded, rare-earth ions are indeed the dominant source for the linewidth peak at low temperatures. 

\section{Conclusions}
We determined the ferromagnetic dispersion and  linewidth of the (110) magnetostatic mode of a polished YIG sphere as a function of temperature and frequency. From this data, we extract the Gilbert damping parameter for temperatures above \SI{100}{\kelvin} and find that it varies linearly with temperature as expected according to the two competing theories of \citet{Kasuya1961} and \citet{Cherepanov1993}. At low temperatures, the temperature dependence of the linewidth measured at constant magnetic field shows a peak that shifts to higher temperatures with increasing frequency. This indicates slowly relaxing impurities as the dominant relaxation mechanism for the magnetostatic modes below \SI{100}{\kelvin}. We model the shift of the peak position with temperature and frequency with values reported for Neodymium impurities\cite{Clarke1965a} in combination with a typical value for the impurity-ion to iron-ion exchange coupling. We find that these parameters can be used to describe the position of the linewidth peak. We thus directly show the implications of (rare earth) impurities as typically present in YIG samples on the dynamic magnetic properties of the ferrimagnetic garnet material. Furthermore, we extract the temperature dependence of the saturation magnetization, the anisotropy field and the $g$-factor.

\section*{Acknowledgements}
The authors thank M. S. Brandt for helping out with the microwave equipment. C.D. and S.O. would like to acknowledge R. Meyer, M. Reich, and B. Wenzel (INNOVENT e.V.) for technical assistance in the YIG crystal growth and sphere preparation.
We gratefully acknowledge funding via the priority program Spin Caloric Transport (spinCAT), (Projects GO 944/4 and GR 1132/18), the priority program SPP 1601 (HU 1896/2-1) and the collaborative research center SFB 631 of the Deutsche Forschungsgemeinschaft.

\bibliography{references}

\end{document}